\documentclass[11pt,a4paper]{article}
\usepackage[margin=23mm]{geometry}
\usepackage[T1]{fontenc}
\usepackage{lmodern,amsmath,amssymb,amsthm,booktabs,array,microtype}
\usepackage[hidelinks]{hyperref}
\hypersetup{pdftitle={Block-exchange descent for insertion},
  pdfsubject={Commuting projections and minimum insertion degree},
  pdfkeywords={insertion, shuffle, trajectories, commuting words, periodicity},
  pdfauthor={Haobo Ma, Wenlin Zhang, Charles E. Hughes}}
\newtheorem{theorem}{Theorem}
\newtheorem{lemma}[theorem]{Lemma}
\newtheorem{proposition}[theorem]{Proposition}

\newtheorem*{periodicconjecture}{Conjecture (Periodic-family degree criterion)}
\newcommand{\U}{\mathrm U}
\newcommand{\V}{\mathrm V}
\newcommand{\Sh}{\operatorname{Sh}}
\newcommand{\eps}{\varepsilon}
\newcommand{\ttw}[1]{\texttt{#1}}
\newcommand{\doi}[1]{\href{https://doi.org/#1}{doi:\nolinkurl{#1}}}
\title{Block-exchange descent for insertion:\\
commuting projections and minimum masks}
\author{Haobo Ma\textsuperscript{1}\quad Wenlin Zhang\textsuperscript{2}\quad
Charles E. Hughes\textsuperscript{3}}
\date{November 2026}
\makeatletter
\renewcommand{\maketitle}{%
\begin{center}
{\LARGE\@title\par}
\vspace{4mm}
\@author\par
\vspace{3mm}
{\small
\textsuperscript{1}ChronoAI Pte Ltd; \texttt{auric@aelf.io}\par
\smallskip
\textsuperscript{2}National University of Singapore; ChronoAI Pte Ltd\par
\texttt{e1327962@u.nus.edu}\par
\smallskip
\textsuperscript{3}Department of Computer Science, University of Central Florida\par
Orlando, FL, USA; \texttt{charles.hughes@ucf.edu}\par}
\vspace{3mm}
\@date\par
\end{center}}
\makeatother
\begin{document}
\maketitle
\begin{abstract}
Singleton No-Gap asks whether every intermediate minimum insertion degree
occurs for a fixed pair of source words.
For fixed nonempty words $u,v$, the minimum insertion degree of a shuffle
is the least number of inserted-source runs among its representations.
We prove that, after a block exchange from a minimum mask of degree $r$,
the minimum over masks with commuting source projections is exactly
$r-1$; a larger decrease occurs precisely when a noncommuting mask has
fewer runs. If either source is unary, every such exchange is exact.
We prove an explicit repair for an infinite periodic degree-three family.
Computer-assisted results establish first-mask
descent through combined length seventeen over every alphabet,
degree-three first-mask descent through length thirty, and a uniform
repair of all degree-three failures through length thirty, including
all further letter identifications. Counterexamples to arbitrary, first,
and last minimum-mask choices explain the need for alternative masks.
Finally, we prove that minimum iteration depths have no gaps for arbitrary
languages. General
block-exchange descent and Singleton No-Gap remain open.
\end{abstract}

\section{Introduction}\label{sec:intro}

Insertion and shuffle are fundamental operations on words and formal
languages. Both combine two sources while preserving the order of the
symbols within each source, but they differ in how freely the symbols
may be interleaved. In this paper, we investigate a natural question
arising from that distinction: how many separate blocks of one word do
we need to insert into another to produce a target word, and which
minimum numbers of blocks can arise as the target word varies?

Let $u$ and $v$ be nonempty words over an alphabet $\Sigma$. We regard
$u$ as the inserted word and $v$ as the base word. A single insertion
places all of $u$ at one position in $v$, without changing the order of
the letters of either word. More generally, a $k$-insertion divides $u$
into at most $k$ nonempty consecutive factors and distributes them into
positions in $v$, again preserving the order of both sources. The
formal definition permits empty factors, so that increasing the bound
$k$ never removes an output.

If we set the insertion position to be the beginning of $v$, the result
is the concatenation of $u$ and $v$, $uv$. If the number of inserted
factors is unrestricted, the resulting words are precisely the shuffles
of $u$ and $v$: all interleavings that preserve the relative order of
the letters originating in each source. Thus, bounded insertion
provides a hierarchy between single-block insertion, such as
concatenation, and unrestricted shuffle.

An important subtlety is that an output word may have several different
representations as a shuffle of the same two sources. When letters
occur in both source words, their origins may be ambiguous.
Consequently, the number of inserted blocks in one representation need
not be the smallest possible number for that output. We call the least
positive number of inserted blocks required to produce an output $w$
its \emph{minimum insertion degree}, denoted $d_{u,v}(w)$.

For example, take $u=00$ and $v=11$. The word $0011$ has minimum degree
$1$, because the two zeros can be inserted together. In contrast,
$0110$ has minimum degree $2$, since its zeros are separated by the base
letters, so two inserted blocks are necessary. Both words are shuffles
of the same sources. This simple example illustrates how minimum
insertion degree measures the amount of separation required between
inserted-source letters. The source mask of $0110$ is
$\ttw{UVVU}$: the two zeros carry label $\U$ and the two ones
carry label $\V$. Its two maximal $\U$-runs correspond to the two
inserted blocks.

For fixed sources $u,v$, consider the set of minimum degrees attained
by all their shuffle outputs:
\[
 D(u,v)=\{d_{u,v}(w):w\in\Sh(u,v)\}.
\]
Degree $1$ is always attained: the concatenation $uv$ is a
shuffle in which the entire inserted word $u$ forms a single block. The
Singleton No-Gap problem asks whether this spectrum always consists of
consecutive positive integers, from $1$ through some maximum. In other
words, if an output requires at least $r$ inserted blocks, must there
also be outputs whose minimum degrees are $1,2,\ldots,r-1$?

This question arises from Hughes's investigation of bounded insertion
and adjacent equality in hierarchies of language
operations~\cite[Section~11]{hughes}. For singleton source languages,
the No-Gap property would imply that the bounded-insertion hierarchy
cannot skip an attainable minimum degree. For general languages,
additional ambiguity arises because an output may be produced by
different source-word pairs. We distinguish this broader insertion-degree
problem from the singleton case considered here.

Our approach studies the representations of a shuffle directly. A
\emph{source mask} labels each output position according to whether its
letter came from $u$ or $v$. The number of maximal runs labeled by the
inserted source equals the number of inserted blocks in that mask.
A \emph{minimum mask} is one attaining the minimum insertion degree.

Suppose a minimum mask has $r>1$ inserted-source runs. Two consecutive
inserted runs are separated by a block of base-source letters.
Exchanging that separating block with either neighboring inserted run
merges the two runs while preserving the order of letters within each
source. The resulting output therefore has a representation with
$r-1$ inserted runs. However, it may admit another representation with
fewer than $r-1$ runs. This is the principal obstacle to obtaining a
degree-$(r-1)$ predecessor.

We ask whether, for every shuffle output of minimum degree $r>1$,
some minimum mask admits a block exchange whose resulting output has
minimum degree exactly $r-1$. An affirmative answer would establish
Singleton No-Gap by repeated descent.

Our principal structural result characterizes the obstruction to an
exact block exchange. We prove that transporting an alternative mask
back through the exchange preserves both source words precisely when
the corresponding projected factors commute. Here two words commute
when concatenating them in either order gives the same word. Among
these commuting masks, the minimum run count after the exchange is
exactly $r-1$; any larger reduction in the output's minimum insertion
degree must arise from a noncommuting mask
(Theorem~\ref{thm:criterion}). This yields an exact obstruction
criterion and a dynamic-programming test.

We then establish that every block exchange from every minimum mask is
exact whenever either source word is unary, meaning that it consists
of repetitions of a single letter (Theorem~\ref{thm:unary}). We also
prove an exact repair theorem for an infinite periodic family of
degree-three obstructions (Theorem~\ref{thm:periodrepair}). Explicit
counterexamples show why an arbitrary minimum mask, or even the
lexicographically first or last minimum mask, cannot always be selected.

Computer-assisted results establish additional bounded cases over
arbitrary alphabets and identify the limitations of several natural
mask-selection rules. Their precise length and degree bounds are stated
in Section~\ref{sec:computation}. Separately, we prove that the spectrum
of minimum iteration depths has no gaps for arbitrary source
languages (Theorem~\ref{thm:depth}), extending the finite-language
result of Hughes~\cite[Section~10]{hughes}. Here iteration depth counts
successive insertion steps; insertion degree counts blocks within one
step.

The paper develops the relevant background, proves the structural and
unary results, examines the counterexamples and periodic repair,
presents the computational certificates, and concludes with the
remaining open questions. The general existence of an exact
block-exchange predecessor and the general Singleton No-Gap conjecture
both remain open.

\section{Related work}\label{sec:related}

Insertion and deletion have long been studied as operations on formal
languages. Haussler's insertion languages~\cite{haussler} provide an
early language-theoretic treatment. Kari studies closure properties of
insertion~\cite{kariinsertion} and deletion~\cite{kari} operations;
Kari and Thierrin~\cite{karithierrin}
study contextual insertion and deletion and their computational power.
Ito, Kari, and Thierrin treat insertion/deletion closure~\cite{itokari}
and shuffle/scattered-deletion closure~\cite{itoshuffle}.
These works concern the languages generated or preserved by the
operations. Our question concerns the least number of inserted blocks
needed for an individual output of a fixed pair of words.

The closest bounded-operation framework is the $n$-insertion of Ito
and Sugiura~\cite{ninsertion}. They study its language-class properties
and decompositions of regular languages under $n$-insertion.
Castiglione and Restivo~\cite{castiglionerestivo} prove that star-free
languages are closed under bounded shuffle. Their bound counts paired
factors from the two sources, whereas ours counts maximal inserted-source
runs. In paired form $x_1y_1\cdots x_ny_n$, with the $x_i$ from the base
and the $y_i$ from the inserted source, there are at most $n$ inserted
runs. Conversely, a mask with $k$ inserted runs has the form
$x_1y_1\cdots x_ky_kx_{k+1}$ and becomes a paired factorization with
$k+1$ pairs by taking $y_{k+1}=\eps$; endpoint factors may be empty.
This relates the bounded operations; their closure theorem concerns
language classes, while our degree spectrum minimizes over representations
of each fixed output.

A bound on inserted blocks is also a restriction on the binary interleaving
trajectory. Mateescu, Rozenberg, and Salomaa~\cite{trajectories}
develop shuffle on trajectories, in which a language of trajectories
specifies allowed interleavings; Domaratzki~\cite{domaratzki} develops
the corresponding deletion operation. In that terminology, masks of
at most $k$ inserted runs form a regular trajectory language. The
minimum degree studied here is the least $k$ for which a given output
has a representation in that language. Distinct trajectories may emit
the same word, and this equality is what makes local descent nontrivial.

Shuffle decomposition is another related inverse problem.
C\^ampeanu, Salomaa, and V\'agv\"olgyi~\cite{decompositions} introduce a
shuffle analogue of the Nerode congruence and study decomposition of
languages into shuffle factors. Their factors are languages to be
recovered, whereas our factors $u,v$ are fixed and the number of blocks
in their interleaving is minimized. Thus $n$-insertion and trajectory
constraints supply the direct connection to interleavings with few
blocks, while language decomposition supplies a related account of
shuffle ambiguity.

The complexity of shuffle depends strongly on which objects are given.
Warmuth and Haussler~\cite{warmuth} study the complexity of iterated
shuffle. Buss and Soltys~\cite{buss} prove that deciding whether a word
is a shuffle of two identical, unspecified words is NP-complete.
For two specified source words, a source-grid dynamic program decides
membership; adding a previous-source label and a cost for starting an
inserted run computes the minimum degree. We use that weighted grid
both for the fixed-exchange formula~\eqref{eq:DP} and for testing
equality partitions in the symbolic search. The computational difficulty
here is the exhaustive search for descent obstructions over all masks
and letter-equality patterns.

The algebraic obstruction is classical word commutation. The
Lyndon--Sch\"utzenberger commutation theorem~\cite{lyndon}
characterizes commuting nonempty words as powers of a common word;
it interprets the projection equations in Lemma~\ref{lem:valid}.
The primitive-root and rotation arguments in
Section~\ref{sec:periodic} belong to the standard combinatorics of
words~\cite{lothaire}. Fine and Wilf's periodicity
theorem~\cite{finewilf} gives a general criterion for two overlapping
periods to force a common smaller period. The rotation argument needed
here has a direct proof, given in Section~\ref{sec:prelim}.

Hughes~\cite[Sections~10--11]{hughes} formulates the two spectrum
questions that motivate this paper. His Theorem~8 establishes the
No-Gap conclusion for iteration depth of finite languages, and his
Theorem~9 gives the unary-source singleton degree spectrum. We prove
the arbitrary-language iteration-depth extension, retain a
self-contained proof of the unary formula, and strengthen its
conclusion to exactness of every minimum-mask exchange. The block
commutation criterion, counterexamples, and periodic repair address
the representation ambiguity in his singleton question.

\section{Preliminaries}\label{sec:prelim}

Let $\Sigma$ be an alphabet, $\Sigma^*$ its free monoid, and $\eps$
the empty word. Write $|x|$ for word length and $\operatorname{alph}(x)$
for the set of letters occurring in $x$. A word is \emph{unary} if
it is a nonempty power of one letter. Positions are indexed from zero
unless explicitly numbered otherwise; $x[i:j]$ denotes a half-open
factor. Concatenation is written by juxtaposition. Spaces in displayed
binary words separate groups of symbols and are not letters.

The shuffle $\Sh(u,v)$ consists of the interleavings of $u$ and $v$
that preserve the order of each source. Throughout
Sections~\ref{sec:criterion}--\ref{sec:computation}, $u,v$ are nonempty
and form a fixed ordered pair: $u$ is the inserted source and $v$ the
base source. A \emph{mask} of $w\in\Sh(u,v)$ labels its positions
$\U$ or $\V$ so that the two projected subsequences are $u$ and $v$.
Write $\rho(M)$ for the number of maximal nonempty $\U$-runs and set
\[
 d(w)=d_{u,v}(w)=\min_{M\text{ a mask of }w}\rho(M),\qquad
 D(u,v)=\{d_{u,v}(w):w\in\Sh(u,v)\}.
\]
A $k$-insertion splits $u=y_1\cdots y_k$ and $v=x_1\cdots x_{k+1}$
and emits $x_1y_1\cdots x_ky_kx_{k+1}$, with empty factors allowed.
For nonempty $u$, its least positive $k$ is exactly $d_{u,v}(w)$:
the nonempty $y_i$ give at most $k$ inserted runs, and the maximal
inserted runs of a mask give such a factorization. Consequently
$1\le d(w)\le\min(|u|,|v|+1)$. Singleton No-Gap asks whether $D(u,v)$
is an initial interval of positive integers.

Let $M$ be a minimum mask of a word of degree $r>1$. A
\emph{block exchange} interchanges a separating maximal $\V$-block
with either neighboring maximal $\U$-run, moving letters and labels
together. Both source orders are preserved, and two $\U$-runs merge.
The resulting word $z$ therefore satisfies $d(z)\le r-1$. The exchange
is \emph{exact} when $d(z)=r-1$. We call $M$ \emph{bad} if none of
its exchanges is exact. Masks are compared lexicographically with
$\U<\V$. Number the separating $\V$-blocks from left to right; $sL$
and $sR$ exchange separator $s$ with its preceding and following
$\U$-runs, respectively. Successor-degree vectors use the order
$1L,1R,2L,2R,\ldots$.

A nonempty word is \emph{primitive} if it is not a proper power of a
shorter word. Formally, a nonempty word $T$ is primitive if, for every
nonempty word $S$ and every positive integer $j$, the equality $T=S^j$
implies $j=1$. In such an equality, $|T|=j|S|$; thus a proper-power
witness necessarily satisfies $2\le j\le |T|$ and
$1\le |S|\le\lfloor |T|/2\rfloor$.
Every nonempty word has a unique primitive root. A primitive word of
length $p$ is fixed by a cyclic rotation through $q$ positions only if
$p\mid q$: rotation invariance makes its letters constant on the
residue classes modulo $\gcd(p,q)$, so it would otherwise be a proper
power of its prefix of that length. This elementary consequence of
primitivity will be used in the periodic repair argument.

\section{The exchange and its exact obstruction}\label{sec:criterion}

\begin{lemma}\label{lem:rotation}
Let $w=XABY$ and $z=XBAY$, with $A,B$ nonempty. Transport any binary
labeling of $z$ with the letters to $w$. Its number of $\U$-runs changes
by at most one.
\end{lemma}
\begin{proof}
Encode $\U$ by $1$ and $\V$ by $0$. Let $\ell$ be the last label of
$X$, or $0$ if $X=\eps$; let $a,\alpha$ be the first and last labels of
$A$, and $b,\beta$ those of $B$; let $t$ be the first label of $Y$, or
$0$ if $Y=\eps$. Internal transitions within the four words are unchanged. If $Y=\eps$,
its end contributes no transition, which is exactly the convention $t=0$.
The run count for $XABY$ minus that for $XBAY$ is
\[
 (1-\ell)(a-b)+(1-\alpha)(b-t)+(1-\beta)(t-a).
\]
If $a=b=t$, this is zero. Otherwise the three differences are a
permutation of $1,-1,0$. The expression is therefore a difference of
two elements of $\{0,1\}$, and belongs to $\{-1,0,1\}$.
\end{proof}

For a mask $N$ of $z=XBAY$, write $A_q,B_q$ for the letter subsequences
in the indicated blocks carrying source label $q\in\{\U,\V\}$.
Call $N$ \emph{commuting} for this exchange if
\begin{equation}\label{eq:commute}
 A_\U B_\U=B_\U A_\U,
 \qquad A_\V B_\V=B_\V A_\V.
\end{equation}
Empty words commute with every word. This definition refers to the
projections of $N$, which may differ from the pure blocks in the selected
minimum mask.

\begin{lemma}\label{lem:valid}
Transporting $N$ back to $w$ gives a mask for the same sources $u,v$
if and only if $N$ is commuting.
\end{lemma}
\begin{proof}
For source $q$, its projection in $z$ has the form
$X_qB_qA_qY_q$, whereas its transported projection in $w$ is
$X_qA_qB_qY_q$. Equality of these two words is equivalent, by left and
right cancellation in the free monoid, to $A_qB_q=B_qA_q$.
Apply this to both sources.
\end{proof}

\begin{theorem}\label{thm:criterion}
Fix a block exchange $E=(X,A,B,Y)$ from a minimum mask of $w=XABY$,
where $d(w)=r>1$ and $z=XBAY$. Put
\[
 c_E(z)=\min_{N\text{ commuting for }E}\rho(N),\qquad
 b_E(z)=\min_{N\text{ noncommuting for }E}\rho(N),
\]
where $N$ ranges over masks of $z$, and an empty minimum is $+\infty$.
Then
\begin{equation}\label{eq:criterion}
 c_E(z)=r-1,\qquad d(z)=\min\{r-1,b_E(z)\}.
\end{equation}
In particular, the exchange is exact if and only if $b_E(z)\ge r-1$.
\end{theorem}
\begin{proof}
If a commuting mask had at most $r-2$ runs, Lemmas~\ref{lem:rotation}
and~\ref{lem:valid} would give a mask of $w$ with at most $r-1$ runs,
contradicting $d(w)=r$. Thus $c_E(z)\ge r-1$. The mask obtained from the
chosen minimum mask has exactly $r-1$ runs and is commuting: each of
the exchanged blocks is pure in one source, so one of its two projections
is empty for each source. Hence $c_E(z)=r-1$. Partitioning all masks of
$z$ into the two classes gives the second identity.
\end{proof}

The equations in Lemma~\ref{lem:valid} have the usual
Lyndon--Sch\"utzenberger interpretation: when both projected factors
are nonempty, they commute precisely when they are powers of a common
primitive word~\cite{lyndon}. For an exchange of adjacent
unequal letters, the commuting masks are exactly those assigning the
two positions to different sources. Different labels give an empty
projection in each source, while identical labels require the two
unequal letters to commute, which is impossible. Thus
Theorem~\ref{thm:criterion} specializes to the different-source and
same-source minimum formula for adjacent-letter pull-back.

\subsection{A source-grid formula}

The commutation predicate can be read from three source-grid states.
In $z=XBAY$, let $t_0=|X|$, $t_1=|XB|$, $t_2=|XBA|$. At these boundaries
let a mask have consumed $i_h$ letters of $u$ and $j_h=t_h-i_h$ letters
of $v$. Its two $\U$-projections are the consecutive factors
\[
 B_\U=u[i_0:i_1],\quad A_\U=u[i_1:i_2],
\]
and similarly $B_\V=v[j_0:j_1]$, $A_\V=v[j_1:j_2]$. Here slices have
zero-based, half-open indices.

Put $m=|u|$ and $n=|v|$. A state is $s=(i,j,p)$, where
$0\le i\le m$, $0\le j\le n$, and $p\in\{\U,\V\}$ is the previous
label. At output position $t$ require $i+j=t$. On an output letter $x$
the only allowed transitions, with their costs, are
\[
\begin{array}{lll}
(i,j,p)\longrightarrow(i+1,j,\U)&\text{if }i<m\text{ and }u[i]=x,
 &\text{cost }\mathbf1_{p=\V},\\
(i,j,p)\longrightarrow(i,j+1,\V)&\text{if }j<n\text{ and }v[j]=x,
 &\text{cost }0.
\end{array}
\]
For any segment $C$, define $T_C(s,s')$ to be the least total cost of
a path reading exactly $C$ from $s$ to $s'$. Compute it by initializing
cost zero at $s$, infinity at every other state, and applying the two
transitions successively to each letter of $C$, retaining the least cost
at each state. A missing path has cost $+\infty$; for an empty segment,
$T_\eps(s,s')=0$ if $s=s'$ and $+\infty$ otherwise. Set
\[
 F_X(s_0)=T_X((0,0,\V),s_0),\qquad
 G_Y(s_2)=\min_{p\in\{\U,\V\}}T_Y(s_2,(m,n,p)).
\]
Thus the initial previous label is $\V$, both sources must be consumed
completely at termination, and an empty prefix or suffix needs no
special convention. With $s_h=(i_h,j_h,p_h)$ at $i_h+j_h=t_h$, the
fixed-exchange obstruction minimum is
\begin{equation}\label{eq:DP}
 b_E(z)=\min_{\substack{s_0,s_1,s_2\\
 \text{at least one source-factor pair does not commute}}}
 \bigl(F_X(s_0)+T_B(s_0,s_1)+T_A(s_1,s_2)+G_Y(s_2)\bigr).
\end{equation}
Splitting a mask at the three boundaries proves one inequality;
concatenating realizing paths with the same boundary states proves the
other. This gives an exact test that accounts for every alternative
mask, including masks with different source consumptions at the block
boundaries.

\subsection{Prefix lower bounds}\label{sec:profiles}

Let $F_z(i,q)$ be the least run count of a mask of the prefix $z$
consuming $i$ letters of $u$ and $|z|-i$ letters of $v$, with last source
$q$. Infeasible states have value $+\infty$, and the initial state has
previous source $\V$. Let $G_x(i,q)$ be the minimum additional run cost
of completing the suffix $x$ from that state. Then
\[
 d_{u,v}(zx)=\min_{i,q}\bigl(F_z(i,q)+G_x(i,q)\bigr).
\]
Splitting a mask at the prefix boundary and concatenating compatible
minimum-cost paths prove the identity. Constraining the labels at two
specified positions gives the corresponding different-source and
same-source minima.

If two words $w,w'$ differ by a swap at positions $i,i+1$, and for some
$i+2\le t<|w|$ one has
\begin{equation}\label{eq:profilebound}
 F_{w'[:t]}(a,q)\ge F_{w[:t]}(a,q)-1
 \quad\text{for every source state }(a,q),
\end{equation}
their common suffix has the same completion cost in both words.
The identity therefore gives $d(w')\ge d(w)-1$.
We call such a swap \emph{profile-certified}. The comparison in
Proposition~\ref{prop:profiles} shows the limits of this sufficient test.

\section{Cases in which every exchange is exact}\label{sec:unary}

\begin{theorem}\label{thm:unary}
If either $u$ or $v$ is unary, every block exchange of every minimum mask
of a word of degree $r>1$ gives an output of degree exactly $r-1$.
\end{theorem}
\begin{proof}
First suppose $u=a^m$, $m>0$. Write
\[
 v=a^{h_0}c_1a^{h_1}\cdots c_sa^{h_s},\qquad c_i\ne a.
\]
Every output has a unique gap description
$a^{h_0+e_0}c_1a^{h_1+e_1}\cdots c_sa^{h_s+e_s}$,
where $e_i\ge0$ and $\sum e_i=m$. Each occupied gap, with $e_i>0$,
requires a $\U$-run; non-$a$ letters are forced to source $\V$.
Putting all inserted $a$'s of each gap together attains the bound, so
\[
 d(w)=\#\{i:0\le i\le s,\ e_i>0\}.
\]
Every minimum mask therefore has one $\U$-run in each occupied gap and
no others. An exchange moves the entire inserted mass of one occupied
gap across the intervening $\V$-block into the adjacent occupied gap.
It empties exactly one occupied gap and hence lowers $d$ by one.

Now suppose $v=a^n$. If $u$ has no non-$a$ letters, $d(w)=1$ and the
assertion is vacuous. Otherwise write
\[
 u=a^{h_0}c_1a^{h_1}\cdots c_sa^{h_s},\qquad s\ge1, c_i\ne a.
\]
The same output-gap description, now with $\sum e_i=n$, gives
\begin{equation}\label{eq:reverse-unary}
 d(w)=1+\#\{i:1\le i<s,\ e_i>0\}.
\end{equation}
Indeed, every non-$a$ letter is forced to $\U$, and every positive
internal gap has a $\V$ between its two bounding non-$a$ letters.
Putting all source-$\U$ copies together in each gap, and putting outer
source-$\V$ copies outside $u$, attains the bound.

Let $k$ be the number of positive internal gaps. These gaps partition
the forced non-$a$ source-$\U$ letters into $k+1$ groups, already
requiring the $k+1$ runs in~\eqref{eq:reverse-unary}. Equality permits
neither an additional all-$a$ $\U$-run nor two $\V$-blocks in one
internal gap: the intervening $\U$-letters would split a forced group
or create an extra run. Thus every $\U$-run contains non-$a$ letters,
and each occupied internal gap contains a single separating
$\V$-block carrying its entire base mass. Exchanging that
block with a neighboring $\U$-run moves the mass into the neighboring
$\V$-block, if present, or into an outer gap. The donor internal gap
becomes empty; the recipient is already an occupied internal gap or is
outer. Exactly one counted gap disappears.
\end{proof}

The first pointwise formula is the gap argument in Hughes's
unary-source spectrum theorem~\cite[Section~11, Theorem~9]{hughes}.
The second formula and both exact-exchange conclusions follow from the
minimum-mask structure proved above.

For completeness, if $s(x)$ denotes the number of non-$a$ letters in
$x$, the formulas give the two spectra
\[
 D(a^m,v)=\{1,\ldots,\min(m,s(v)+1)\},\qquad
 D(u,a^n)=\{1,\ldots,1+\min(n,\max(0,s(u)-1))\}.
\]
In the first case distribute the $m$ inserted letters among any chosen
number of gaps; in the second occupy any chosen number of internal gaps
and put the remaining base letters at an outer end. These constructions
attain every integer up to the stated bounds.

If the source alphabets are disjoint, every mask is uniquely determined
by its letters, so the intended exchanged mask is the only mask and the
exchange is exact. Also, for arbitrary sources and $r=2$, every exchange
is exact: it has a one-run mask, and nonempty $u$ forces $d\ge1$.

\section{Why the minimum mask and separator matter}\label{sec:examples}

The following example rules out choosing an arbitrary minimum mask.
\begin{proposition}\label{prop:badmask}
For
\[
 u=\ttw{01010},\quad v=\ttw{01010101},\quad
 w=\ttw{0010\hspace{0.25em}1100\hspace{0.25em}1101\hspace{0.25em}0},
\]
$d(w)=4$, and the minimum mask $M=\ttw{UVVVVUVUUVVVU}$ has all six block
exchanges of degree $2$.
\end{proposition}
\begin{proof}
Use one-based positions throughout this proof. Both sources alternate.
Each of the equal-letter pairs at positions
$1,2$; $5,6$; $7,8$; and $9,10$ must therefore split between the two
sources. The final $0$ must be $\U$, since $v$ ends in $1$. These choices
use all five $\U$ positions, so every other position is $\V$.
Conversely, every choice of one $\U$ in each of those four pairs,
followed by the final $\U$, gives the correct two sources. This gives
exactly the sixteen masks in Appendix~\ref{app:masks}, Table~\ref{tab:masks}.

The first and last $\U$ positions are isolated. Among the other three,
at most one adjacent pair can merge: the middle one is at position $7$
or $8$, and cannot join both its neighbors. Thus every mask has at least
four runs, and $M$ attains four. Its six exchanges have the four distinct outputs
in Table~\ref{tab:successors}, with multiplicities $2,1,2,1$.
The displayed masks prove degree at most two. A degree-one mask would
select a contiguous occurrence of $u$ whose deletion leaves $v$.
The last column lists the remainders for all such occurrences; none is
$\ttw{01010101}$. Thus all four outputs have degree two.
\end{proof}

\begin{table}[ht]
\centering\small
\caption{The four successor words, two-run witnesses, and all remainders
after deleting a contiguous copy of $u$.}\label{tab:successors}
\begin{tabular}{@{}lll@{}}
\toprule Output & Two-run mask & Remainders\\\midrule
\ttw{0101\hspace{0.25em}0100\hspace{0.25em}1101\hspace{0.25em}0}&\ttw{VVVVVVUVVUUUU}&\ttw{10011010}, \ttw{01011010}\\
\ttw{0010\hspace{0.25em}1010\hspace{0.25em}1101\hspace{0.25em}0}&\ttw{UVVVVVVVVUUUU}&\ttw{01011010}, \ttw{00111010}\\
\ttw{0010\hspace{0.25em}1101\hspace{0.25em}0101\hspace{0.25em}0}&\ttw{VUUUUVVVVVVVU}&\ttw{00101110}, \ttw{00101101}\\
\ttw{0010\hspace{0.25em}1100\hspace{0.25em}1010\hspace{0.25em}1}&\ttw{VUUUUVUVVVVVV}&\ttw{00101101}\\
\bottomrule
\end{tabular}
\end{table}

This example has eight minimum masks. Its first minimum mask,
$\ttw{UVVVUVVUUVVVU}$ with $\U<\V$, has an exchange to
$\ttw{0100\hspace{0.25em}1100\hspace{0.25em}1101\hspace{0.25em}0}$ of degree $3$. Thus
Proposition~\ref{prop:badmask} does not refute existence over all
minimum masks.

\begin{samepage}
\begin{proposition}\label{prop:family}
Let $A\ne B$ be letters, $T=ABA$, and $k,\ell\ge2$. Put
\[
 u=ABABA,\quad v=T^{k+\ell},\quad
 w=AB\,T^k\,A\,T^\ell\,BA.
\]
The word $w$ has exactly twelve masks, each with three $\U$-runs. The mask
\[
 M=\U^2\V^{3k}\U\V^{3\ell}\U^2
\]
has both exchanges at its first separator equal to $vu$, and both at
its second separator equal to $uv$, all of degree one. Its first
minimum mask has an exact degree-two exchange.
\end{proposition}
\end{samepage}
\begin{proof}
Write $N=k+\ell$ and number the $N+2$ occurrences of $B$ in $w$ from
$0$ to $N+1$. The numbers of $A$'s in the initial, internal, and final
gaps are
\[
 (1,\ 1,\ \underbrace{2,\ldots,2}_{k-1},\ 3,
 \underbrace{2,\ldots,2}_{\ell-1},\ 1,\ 1).
\]
The source $v=T^N$ requires two $A$'s between consecutive $B$'s.
Hence at least one of $B_0,B_1$ must be $\U$, and at least one of
$B_N,B_{N+1}$ must be $\U$: retaining both members of either pair would
leave at most one intervening $A$. The source $u$ has exactly two $B$'s,
so it selects one from each pair and no other $B$.

Since $k,\ell\ge2$, both $B$'s bounding the central gap of three $A$'s
are $\V$. Exactly one of those three $A$'s must therefore be $\U$.
It is the middle $A$ of the source $u$. If the first $\U$-labeled $B$
is $B_0$, its preceding $\U$-labeled $A$ is the initial $A$; if it is
$B_1$, that $A$ is the single $A$ between $B_0$ and $B_1$, since the
initial $A$ must remain in $v$. The analogous statement at the right
end selects the following $A$ next to the last $\U$-labeled $B$.
Thus there are exactly $2\cdot3\cdot2=12$ masks. Each has a first
$AB$ run, an isolated central $A$ run, and a last $BA$ run, and all
choices give the two specified sources.

For $M$, the first pair of rotations gives
$T^{N+1}BA=T^N(ABABA)=vu$, since $AB\,A=A\,BA=T$.
The second pair gives $ABT^{N+1}=(ABABA)T^N=uv$.

With $\U<\V$, the first minimum mask is
\[
 M_0=\U^2\V^{3k-1}\U\V^{3\ell-1}\U^2\V^2.
\]
Exchange its first $\U$-run $AB$ with its first separating block
$T^{k-1}AB$. The resulting word is
\[
 z=T^{k-1}ABABAA\,T^\ell BA,
\]
and the transported mask has two runs. A degree-one mask would delete
a contiguous $ABABA$ from $z$. Its two $B$'s would have to be separated
by a single $A$. There are exactly two such internal gaps in $z$.
Deleting the occurrence at the first merges $A$-gaps of lengths $2,1,3$
and removes three $A$'s, leaving an internal gap of length $3$.
Deleting the occurrence at the last leaves the central gap of length
$3$ untouched. Neither remainder is $T^N$, whose internal gaps all have
length $2$. Thus $d(z)=2$.
\end{proof}

For $k=\ell=2$, this gives $u=\ttw{ABABA}$,
$v=\ttw{ABAABAABAABA}$ and $w=\ttw{ABABAABAAABAABABA}$, of total length
seventeen. Proposition~\ref{prop:symbolic} proves that this is the first possible length for an arbitrary-mask failure
at degree three, over every alphabet.
Appending $B^q$ to its base and output forces a final $\V^q$, since
$u$ ends in $A$. Thus an arbitrary minimum mask with no exact exchange exists at every
binary length at least seventeen, already at degree three. Proposition~
\ref{prop:badmask} similarly gives degree-four examples at every binary
length at least thirteen after the same forced-$\V$ suffix padding.

Simple separator rules also fail, even after choosing a first minimum
mask. Table~\ref{tab:rules} gives counterexamples. In each row the
original degree is three, both directions at every selected separator
give degree one, and another separator gives an exact degree-two
predecessor. ``Shortest'' and ``longest'' refer to the separating
$\V$-block, with all ties allowed.
In Table~\ref{tab:rules}, a
complete position check gives the first minimum masks and the selected
degrees $(d_L,d_R)$. Each displayed exact successor has the indicated
two-run mask and contains no contiguous copy of $u$, proving degree two.

\begin{table}[ht]
\centering\footnotesize
\caption{Failed separator choices in the first minimum mask.
Source identity is fixed in each row; $\U<\V$.}\label{tab:rules}
\begin{tabular}{@{}lp{36mm}p{42mm}p{43mm}@{}}
\toprule Rule & Sources and output & First mask; selected degrees & Exact move; output and mask\\\midrule
First &\shortstack[l]{$u=\ttw{011001}$\\$v=\ttw{1001}$\\$w=\ttw{0101011001}$}
&\shortstack[l]{\ttw{UVVUVVUUUU}\\$s=1$, length $2$\\$(d_L,d_R)=(1,1)$}
&\shortstack[l]{$2L$, degree $2$\\\ttw{0100111001}\\\ttw{UVVVVUUUUU}}\\[4pt]
Last &\shortstack[l]{$u=\ttw{01201}$\\$v=\ttw{012}$\\$w=\ttw{01201021}$}
&\shortstack[l]{\ttw{UUUVVUVU}\\$s=2$, length $1$\\$(d_L,d_R)=(1,1)$}
&\shortstack[l]{$1L$, degree $2$\\\ttw{01012021}\\\ttw{VVUUUUVU}}\\[4pt]
Shortest &\shortstack[l]{$u=\ttw{0011001}$\\$v=\ttw{0011}$\\$w=\ttw{00110010101}$}
&\shortstack[l]{\ttw{UUUUVVVUVUU}\\$s=2$, length $1$\\$(d_L,d_R)=(1,1)$}
&\shortstack[l]{$1L$, degree $2$\\\ttw{00100110101}\\\ttw{VVVUUUUUVUU}}\\[4pt]
Longest &\shortstack[l]{$u=\ttw{0101010}$\\$v=\ttw{010}$\\$w=\ttw{0100110010}$}
&\shortstack[l]{\ttw{UUUVUVVUUU}\\$s=2$, length $2$\\$(d_L,d_R)=(1,1)$}
&\shortstack[l]{$1L$, degree $2$\\\ttw{0010110010}\\\ttw{VUUUUVVUUU}}\\
\bottomrule
\end{tabular}
\end{table}

\begin{samepage}
\begin{proposition}\label{prop:firstfailure}
For
\[
 u=\ttw{ABAABAABAAB},\quad v=\ttw{ABABAAB},\quad
 w=\ttw{ABABAAABABABAAABAB},
\]
the first minimum mask, with $\U<\V$, is
$M_0=\ttw{UUVVVUUUUVUUUUVUVV}$. It is a minimum mask for a word of
degree four, but all six
block exchanges give degree two. Another minimum mask has an exact
degree-three exchange.
\end{proposition}
\end{samepage}
\begin{proof}
Number the seven $B$'s of $w$ from $0$ to $6$. The $A$-gap lengths are
$(1,1,3,1,1,3,1,0)$. Consecutive $B$'s in $u$ require two intervening
$A$'s. Therefore $u$ can take at most one of $B_0,B_1$, at most two
of $B_2,B_3,B_4$, and at most one of $B_5,B_6$. It needs four $B$'s,
so it takes one from each outer pair and two in the middle.
The pairs $B_2,B_3$ and $B_3,B_4$ have only one intervening $A$,
whereas consecutive $B$'s in $u$ need two. The middle pair is therefore
exactly $B_2,B_4$.
The remaining three $B$'s belong to $v$.

The first source-$\V$ gap between $B$'s needs one $A$, and the second
needs two. Assigning the $A$'s accordingly gives exactly the following
six prefix masks of the first ten positions, and six suffix masks of
the last eight positions:
\[
\begin{array}{ll}
\text{prefix}&
\ttw{UUVVUUVUUV},\ \ttw{UUVVUVUUUV},\ \ttw{UUVVVUUUUV},\\
&\ttw{VVUUUUVUUV},\ \ttw{VVUUUVUUUV},\ \ttw{VVUUVUUUUV};\\[2pt]
\text{suffix}&
\ttw{UUUUVUVV},\ \ttw{UUUVUUVV},\ \ttw{UUUVVVUU},\\
&\ttw{UUVUUUVV},\ \ttw{UUVUVVUU},\ \ttw{UUVVUVUU}.
\end{array}
\]
Every prefix projects to $(\ttw{ABAABA},\ttw{ABAB})$, every suffix
to $(\ttw{ABAAB},\ttw{AAB})$, and every combination is valid.
Each prefix ends in $\V$, so the run counts add. Four prefixes and
four suffixes have two runs; the other two in each list have three.
Thus there are exactly $36$ masks, with $16,16,4$ masks having
$4,5,6$ runs, respectively. The first two-run prefix and first two-run
suffix give $M_0$.

The six exchanges of $M_0$ have five distinct outputs, listed in
Table~\ref{tab:firstfailure}; the first occurs twice. The displayed
two-run masks give degree at most two. None of these words contains
a contiguous copy of $u$, so none has degree one. Finally,
$\ttw{UUVVVUUUUVUUUVUUVV}$ is another minimum mask; exchanging its
second separator with the following $\U$-run (the $2R$ exchange) gives
$\ttw{ABABAAABAABABAABAB}$. The transported mask
$\ttw{UUVVVUUUUUUUVVUUVV}$ has three runs. This word's internal
$A$-gaps between successive $B$'s have lengths $(1,3,2,1,2,1)$.
Selecting the four $B$'s of $u$ in at most two runs would require
at least two pairs of successive $B$'s within a run, each separated
by exactly two $A$'s. The two gaps of length two would therefore
both have to be selected. But their four bounding $B$'s have only
one $A$ between the middle pair, whereas $u$ requires two. Thus
two runs are impossible and the degree is exactly three.
\end{proof}

The first-mask failure occurs at every binary length at least eighteen.
Append $A^q$ to $v$ and $w$ in Proposition~\ref{prop:firstfailure}.
Every appended position is forced to $\V$: if any were $\U$, the last
$\U$-letter would be $A$, whereas $u$ ends in $B$. Thus all masks,
their run counts, and their lexicographic order are unchanged apart
from the final $\V^q$. The same applies to every exchanged word.

\begin{table}[ht]
\centering\small
\caption{The five degree-two outputs from the first minimum mask in
Proposition~\ref{prop:firstfailure}.}\label{tab:firstfailure}
\begin{tabular}{@{}ll@{}}
\toprule Output & Two-run mask\\\midrule
\ttw{ABAABAABABABAAABAB}&\ttw{UUUUUUUUVVVVVUUUVV}\\
\ttw{ABABABAABAABAAABAB}&\ttw{VVVVUUUUUUUUUUVUVV}\\
\ttw{ABABAAABAABAABABAB}&\ttw{VVVVVVUUUUUUUUUVUU}\\
\ttw{ABABAAABABAABAABAB}&\ttw{VVUUUUVVVUUUUUUUVV}\\
\ttw{ABABAAABABABAABAAB}&\ttw{VVVVVVUUUVUUUUUUUU}\\
\bottomrule
\end{tabular}
\end{table}

\begin{proposition}\label{prop:lastfailure}
For
\[
 u=\ttw{ABABABBABAB},\quad v=\ttw{ABABABBAB},\quad
 w=\ttw{ABABABBABAABBABABABB},
\]
$d(w)=4$. The last minimum mask, $\ttw{UUUUUUUVVVUUVVVUVVVU}$,
has all six block exchanges of degree two. The first minimum mask,
$\ttw{UUUUUUUVVVUUVVVUVVUV}$, has successor-degree vector
$(2,2,2,2,3,2)$, so its $3L$ exchange is exact.
\end{proposition}
\begin{proof}
The exhaustive position test gives exactly the eight masks in
Appendix~\ref{app:masks}, Table~\ref{tab:lastcert}: six have
five runs and two have four. To check completeness of this finite list,
test each choice of $|u|$ positions in $w$ against $u$, and test the
complementary positions against $v$, retaining precisely the matches.
The last minimum mask is $\ttw{UUUUUUUVVVUUVVVUVVVU}$; all six of
its successor degrees are $(2,2,2,2,2,2)$. Their five distinct outputs and
two-run masks appear in Table~\ref{tab:lastfailure}. Four outputs
contain no contiguous $u$; the remaining output has the sole
contiguous-$u$ deletion remainder $\ttw{ABABABABB}$, different from
$v$. Thus the displayed upper bounds are exact.
The first minimum mask, $\ttw{UUUUUUUVVVUUVVVUVVUV}$, has successor
degrees $(2,2,2,2,3,2)$. Its sole exact move is $3L$, to
$z_*=\ttw{ABABABBABAABBABBAABB}$. The same appendix table also
lists all sixteen masks of $z_*$, obtained by the same finite position
test. None has fewer than three runs, and
$\ttw{UUUUUUUVVVUUVVVVVUUV}$ has three. This proves $d(z_*)=3$.
\end{proof}
Proposition~\ref{prop:symbolic} establishes length twenty as the
smallest degree-four last-mask failure over every alphabet.
Neither extremal minimum mask alone is a universal
choice.

\begin{table}[ht]
\centering\small
\caption{Degree-two exchanges of the last minimum mask in the
length-twenty example. The second output occurs twice.}\label{tab:lastfailure}
\begin{tabular}{@{}ll@{}}
\toprule Output & Two-run mask\\\midrule
\ttw{ABAABABABBABBABABABB}&\ttw{VVVUUUUUUVVVVVUUUUUV}\\
\ttw{ABABABBABABABABABABB}&\ttw{VVVVVVVUUUUUUVUUUUUV}\\
\ttw{ABABABBABAABABABBABB}&\ttw{UUUUUUUUUUVVVVVVVVUV}\\
\ttw{ABABABBABAABBABBABAB}&\ttw{VVVVVVVUUUVVUUUUUUUU}\\
\ttw{ABABABBABAABBABABBAB}&\ttw{UUUUUUUUUUVUVVVVVVVV}\\
\bottomrule
\end{tabular}
\end{table}

Even allowing both end separators of the first minimum mask is
insufficient. Take
\[
 u=\ttw{01201Y34534},\quad v=\ttw{201XZ345},\quad
 w=\ttw{02101201XYZ34534354}.
\]
Its first minimum mask is $\ttw{UVUVVUUUVUVUUUVVUVU}$, a minimum mask
for a word of degree seven. The four moves at the first and last
separators have degree five. To see the end failures,
split the word at the fresh markers $XYZ$, whose labels are forced
$\V\U\V$. The left gadget $(\ttw{01201},\ttw{201},\ttw{02101201})$
has degree three and both first-separator moves give degree one.
The right gadget $(\ttw{34534},\ttw{345},\ttw{34534354})$
has degree three and both last-separator moves give degree one.
Their alphabets are disjoint, and the markers isolate one extra
$\U$-run. The global degree is therefore the sum of the two local
degrees plus one, both before and after these end moves.

\subsection{Label shifts and a further counterexample}

The first minimum mask satisfies useful necessary conditions. Suppose
one of its separating $\V$-blocks is $P$ and its following $\U$-run
is $B$. If $P=P_0B$ with $P_0\ne\eps$, the two adjacent copies of $B$
can exchange source labels: $\U a\,\V(P_0B)\,\U B$ becomes
$\U a\,\V P_0\,\U B\,\V B$. If $B=PB_0$ with $B_0\ne\eps$, the
two copies of $P$ can exchange labels, changing
$\U a\,\V P\,\U(PB_0)$ to
$\U(aP)\,\V P\,\U B_0$.
Both changes preserve the source words and the number of runs, and
give an earlier mask. The first minimum mask therefore admits neither
arrangement. These are shifts within one output word, not descent
exchanges to a different output.

These shifts do not characterize the obstruction. For example,
\[
 u=\ttw{ABBAABBA},\quad v=\ttw{ABBABABBA},\quad
 w=\ttw{ABBABABABABABABBA}
\]
has exactly the eight masks listed in Appendix~\ref{app:masks},
Table~\ref{tab:packedcert},
all with four $\U$-runs. The selected minimum mask
$\ttw{VVVUUVUUVUUVUUVVV}$ has successor degrees $(2,2,2,2,2,2)$;
Table~\ref{tab:packedmoves} gives the six outputs and two-run witnesses.
For the first output the sole contiguous-$u$ deletion remainder is
$\ttw{BABABABBA}$, and for the last it is $\ttw{ABBABABAB}$; neither
is $v$. The other four outputs contain no contiguous $u$. Thus all
six degrees are exactly two.

In the selected mask the three internal $(P,B)$ pairs of a separating
$\V$-block and its following $\U$-run are $(A,BA)$, $(B,AB)$, and
$(A,BA)$. Each separator is shorter than its following run and cannot
have that run as a proper suffix; neither $BA$ begins with $A$ nor
$AB$ with $B$.
At the leading $\V$-block the pair is $ABB,AB$: $ABB$ does not end
with $AB$, and $AB$ is shorter. Hence neither whole-factor shift is
available, including at the leading block.

The first minimum mask is $\ttw{UUUVVUVUUVUUVVVVV}$; its $1L$ move
gives $z_\dagger=\ttw{ABABBABABABABABBA}$. Table~\ref{tab:packedcert}
lists all twenty-one masks of $z_\dagger$. The minimum run count in
that list is three, attained for example by
$\ttw{VVUUUUVUUVUUVVVVV}$. The same exhaustive position test as above
produces both complete lists, proving $d(z_\dagger)=3$. 

\begin{table}[ht]
\centering\small
\caption{The six degree-two successors of
$\ttw{VVVUUVUUVUUVUUVVV}$ in the packed example.}\label{tab:packedmoves}
\begin{tabular}{@{}cllc@{}}
\toprule Move & Output & Two-run mask & Degree\\\midrule
1L & \ttw{ABBAABBABABABABBA} & \ttw{UUUUUUUVVUVVVVVVV} & 2\\
1R & \ttw{ABBABBAABABABABBA} & \ttw{VVVUUUUUUVUUVVVVV} & 2\\
2L & \ttw{ABBABABBAABABABBA} & \ttw{VVVVVUUUUUUVUUVVV} & 2\\
2R & \ttw{ABBABABAABBABABBA} & \ttw{VVVUUVUUUUUUVVVVV} & 2\\
3L & \ttw{ABBABABABAABBABBA} & \ttw{VVVUUVVVUUUUUUVVV} & 2\\
3R & \ttw{ABBABABABABBAABBA} & \ttw{VVVVVVVUVVUUUUUUU} & 2\\
\bottomrule
\end{tabular}
\end{table}

\section{A periodic repair at degree three}\label{sec:periodic}

The preceding counterexamples show that a minimum mask can have no
exact block exchange even when another minimum mask does. We now
study a mechanism for changing the source labels without changing
the output or its minimum run count. A common periodic structure
forces such a shift; for an explicitly chosen infinite family, we also prove
that the shifted mask admits an exact descent exchange.

\subsection{A label shift under a common-period hypothesis}

Write a three-run mask as
\[
 w=L\,a\,P\,b\,Q\,c\,R,
 \qquad u=abc,\quad v=LPQR,
\]
where $a,b,c,P,Q$ are nonempty and the outer blocks $L,R$ may be empty.
The bounded classification in Proposition~\ref{prop:boundedrepair} gives
$P,Q,ab,bc$ as powers of one primitive word $T$ for every bad
degree-three mask in its domain. We first examine the consequence
of this common-period hypothesis.

Suppose now that $d(w)=3$ and $P,Q,ab,bc$ are powers of one
primitive word $T$. This common-period hypothesis forces a label shift. Put
$p=|T|$ and $q=|b|$. If $q\ge p$, the first $p$ letters of $b$ are $T$
from $bc$, and are the rotation $\operatorname{rot}_{-q\bmod p}(T)$
from $ab$: $T=\operatorname{rot}_{-q\bmod p}(T)$. Primitivity forces
$p\mid q$, and then $a,b,c,P,Q$ are powers of $T$.
Selecting $abc$ contiguously in the local periodic block would give a
one-run mask of $w$, even with the outer contexts, a contradiction.
Thus $|b|<|T|$. Hence $|P|>|b|$, and $P=P_0b$
with $P_0\ne\eps$. Relabeling the adjacent identical copies of $b$ gives
\begin{equation}\label{eq:periodshift}
 \U a\,\V(P_0b)\,\U b\,\V Q\,\U c
 \quad\longmapsto\quad
 \U a\,\V P_0\,\U b\,\V(bQ)\,\U c.
\end{equation}
The two source words and the three-run count are unchanged, and the new
mask is earlier under $\U<\V$. Existence of this shift alone does not
prove that one of its exchanges is exact. The following result supplies
the missing degree calculation for an infinite family.

\subsection{Exact repair for an infinite family}

\begin{theorem}\label{thm:periodrepair}
Let $A\ne B$ be letters, put $T=ABA$, and let $h,j,k,\ell$ be positive
integers. Set
\[
 a=T^{h-1}AB,\quad b=A,\quad c=BA\,T^{j-1},\qquad
 u=abc,\quad v=T^{k+\ell},\quad w=aT^k A T^\ell c.
\]
The mask $M=\U^{|a|}\V^{3k}\U\V^{3\ell}\U^{|c|}$ has three runs;
its four exchanges give $vu,vu,uv,uv$, all of degree one. Shifting the
middle $A$ to the identical final $A$ of $T^k$ gives
\[
 N=\U^{|a|}\V^{3k-1}\U\V^{3\ell+1}\U^{|c|}.
\]
The $1L$ exchange of $N$ gives
\begin{equation}\label{eq:periodexact}
 z=T^{k-1}AB\,T^h A\,T^\ell BA\,T^{j-1},\qquad d_{u,v}(z)=2.
\end{equation}
Consequently, whenever $d_{u,v}(w)=3$, $N$ is an alternative minimum
mask with a specified exact exchange.
\end{theorem}

\begin{proof}
The identities $ab=T^h$ and $bc=T^j$ give the four exchanged words.
For example, both exchanges at the first separator give
$T^{k+h+\ell}c=T^{k+\ell}u=vu$; those at the second give $uv$.
The shift preserves the source projections, and its $1L$ exchange
has the word in~\eqref{eq:periodexact} and a two-run transported mask.
It remains to rule out degree one.

Number the $K=h+j+k+\ell$ copies of $B$ in $z$ from zero. Its initial
and final $A$-gaps have length one. All internal gaps have length two,
except the gaps after $B_{k-1}$, $B_{k+h-1}$, and $B_{k+h+\ell-1}$,
whose lengths are respectively $1,3,1$. A contiguous copy of $u$ uses
$h+j$ successive $B$'s and has exactly one internal gap of length one,
after its $h$th $B$; all its other internal gaps have length two.

If this short gap matches the first short gap of $z$, the copy must
start at $B_{k-h}$ and end at $B_{k+j-1}$. A negative initial index is
impossible. If $j>h$, the copy contains the gap of length three and
cannot be $u$. If $j<h$, deleting it leaves that gap untouched. If
$j=h$, the deletion ends beside that gap and removes one of its three
$A$'s. If $k>h$, the joined internal gap has length
$(2-1)+(3-1)=3$; if $k=h$, the initial gap has length
$(1-1)+(3-1)=2$. The case $k<h$ was excluded by the negative start.
None is compatible with $v=T^{k+\ell}$.

If the short gap matches the last short gap of $z$, the copy starts
at $B_{k+\ell}$ and ends at $B_{K-1}$. When $\ell<h$ it contains the
three-letter gap; when $\ell>h$ the deletion leaves that gap untouched;
when $\ell=h$ the deletion leaves a final gap of length two.
Again the remainder cannot be $v$. These are all possible contiguous
copies of $u$, so no one-run mask exists and $d(z)=2$.
\end{proof}

Charles E. Hughes proposed the following characterization of the
hypothesis in Theorem~\ref{thm:periodrepair}.
\begin{periodicconjecture}
For positive integers $h,j,k,\ell$, with $u,v,w$ as in
Theorem~\ref{thm:periodrepair},
\[
 d_{u,v}(w)=3
 \quad\Longleftrightarrow\quad
 k>h,\quad k\ge j,\quad \ell>j,\quad \ell\ge h.
\]
\end{periodicconjecture}
An independent source-grid calculation over $1\le h,j,k,\ell\le7$
(all $2{,}401$ tuples) found $371$ degree-three instances and no exception
to this criterion. The same calculation agrees with complete mask
enumeration on the box $1\le h,j,k,\ell\le4$. These finite checks support
the conjecture but do not prove it for all positive parameters;
Theorem~\ref{thm:periodrepair} therefore retains its explicit condition
$d_{u,v}(w)=3$.

For $h=j=1$ and $k,\ell\ge2$, Proposition~\ref{prop:family} proves
$d(w)=3$, so this repairs each bad mask in that family without choosing
an extremal mask. The same construction allows $B$ to be replaced by
any nonempty word $S$ containing no $A$. Every $S$-block is pure in a
valid mask. More precisely, every maximal non-$A$ block in each of
$u,v,w$, and the exchanged word $z$ is exactly one copy of $S$:
copies of $S$ are separated by an $A$, and the source order forces both
sources to enter each such block at a copy boundary. Splitting that
block between the sources would leave both copies unfinished at the
following $A$, which neither source could then read. Contracting the
pure $S$-blocks to $B$ therefore
preserves masks and their run counts. In particular, $T=ABCA$ is covered
by taking $S=BC$, where $A,B,C$ are distinct letters.

\section{Computational results}\label{sec:computation}

The computations below enumerate shuffle masks and letter-equality
patterns. The software and data are available as release v1.0 of
\emph{Insertion-degree descent}~\cite{software}; its README gives the
commands, input files, certificates, and execution records. The
propositions state the exact domains of the finite conclusions, and
the following arguments establish completeness of their reductions.
Table~\ref{tab:evidenceoverview} summarizes the scope and the kind of
evidence supporting the main computational statements. A completed
exhaustive search, a certificate check, and an independently regenerated
search are distinct forms of evidence; their respective roles are
specified below.

\begin{table}[htbp]
\centering\small
\caption{Guide to the computational results. Length means
$L=|u|+|v|$; all-alphabet claims use symbolic equality-pattern
enumeration.}\label{tab:evidenceoverview}
\begin{tabular}{@{}p{0.18\linewidth}p{0.39\linewidth}>{\raggedright\arraybackslash}p{0.34\linewidth}@{}}
\toprule
Result & Scope & Evidence \\
\midrule
Prop.~\ref{prop:numerical} & Alphabet sizes at most 2, 3, 4;
$L\le16,12,10$, respectively & Exhaustive numerical enumeration \\
Prop.~\ref{prop:boundedrepair} & Degree-three bad masks,
$L\le30$, all alphabets & Symbolic search and independently checked
certificates; regeneration through $L=20$ \\
Prop.~\ref{prop:symbolic} & All degrees through $L=17$,
all alphabets; additional degree-specific bounds & Symbolic exhaustive
search with completeness reduction \\
Prop.~\ref{prop:profiles} & 168 failed candidate adjacent swaps
in the stated finite domain & Direct enumeration and prefix-profile
certification tests \\
\bottomrule
\end{tabular}
\end{table}

\subsection{Numerical enumeration}\label{sec:numerical}

For each ordered source pair, enumerate all binary masks with $|u|$
$\U$-positions and $|v|$ $\V$-positions, emit their words, and group
equal outputs. The minimum run count is taken after grouping, and the
first and last minimizers are retained. For each selected mask of
degree greater than one, form every block exchange and look up its
output degree in the same table. Every mask has the prescribed
projections by construction, and every representation of an output
occurs in its group. Since an exchange preserves the sources, its
successor is present in that table. This proves completeness for each
finite source-pair domain.

For alphabet bounds, restricted-growth strings for $uv$ represent each
letter-equality pattern once: scanning from left to right, give the
first new letter the name $0$, the next $1$, and so on. Renaming letters
preserves all masks and degrees. Thus enumeration of these patterns
with at most $s$ classes covers every alphabet of size at most $s$.

\begin{proposition}[Computer-assisted]\label{prop:numerical}
For nonempty ordered source pairs in each domain in
Table~\ref{tab:domains}, every output of degree $r>1$ has an exact
exchange from its first minimum mask and an exact exchange from its
last minimum mask. In particular, Singleton No-Gap holds in each domain.
\end{proposition}

\begin{table}[ht]
\centering
\caption{Exhaustive alphabet-bounded domains of
Proposition~\ref{prop:numerical}; $L=|u|+|v|$.}\label{tab:domains}
\begin{tabular}{@{}lc@{}}
\toprule Alphabet size at most & Combined length\\\midrule
Two & $L\le16$\\
Three & $L\le12$\\
Four & $L\le10$\\
\bottomrule
\end{tabular}
\end{table}

The enumerations just described test both extremal masks in these
domains. Repeated exact descent from degree $r$ gives degrees
$r-1,r-2,\ldots,1$ with the same ordered sources, proving the stated
No-Gap consequence. All words in that induction have the same combined
length. The source-pair and output totals are recorded in the software
README.

\subsection{Symbolic enumeration and completeness}\label{sec:symbolic}

Fix length $L$, degree parameter $r\ge3$, and a source-position mask
$M$ of length $L$ with $\rho(M)=r$. Write $m=|M|_\U$ and $n=L-m$,
requiring $m,n>0$. The symbolic search enumerates every such $M$.
Give the $m+n$ positions of the concatenated sources $uv$ distinct
variables $0,\ldots,L-1$. For a mask $H$ with $m$ $\U$-positions
and $n$ $\V$-positions, let $I(H)$ be the variable sequence obtained
by reading each source in order. Then $W_0=I(M)$ is the original
output, and exchanging its pure blocks gives variable sequences
$W_1,\ldots,W_{2(r-1)}$ in the move order defined in Section~\ref{sec:prelim}.

The lower-mask list $\mathcal H$ contains \emph{every} length-$L$
mask with exactly $m$ $\U$-positions, exactly $n$ $\V$-positions, and
$1\le\rho(H)\le r-2$. To propose $H$ as a lower witness for move
$e$, impose all $L$ equalities
\[
 W_e[t]\sim I(H)[t],\qquad 0\le t<L.
\]
These equalities test both source projections, including every consumed
position; equality of the whole emitted words is equivalent to $H$
being a mask of that successor. Union-find represents the least
equivalence relation containing the accumulated equalities. Its
classes give the most general source letters satisfying them.
Further identifications can introduce masks but cannot remove a valid mask.

For fixed $(L,r,M,\mathrm{mode})$, the recursive search is:
\begin{quote}\small
\begin{tabbing}
\quad\=\quad\=\quad\=\kill
$\mathrm{Search}(h,\Pi)$, initially $h=0$ and the discrete partition $\Pi$:\\
\>If $h=2(r-1)$, record a surviving terminal partition and return.\\
\>For every $H\in\mathcal H$:\\
\>\>$\Pi'\gets\Pi\vee\{W_{h+1}[t]\sim I(H)[t]:0\le t<L\}$.\\
\>\>In memoized mode, skip if $(h+1,\operatorname{norm}(\Pi'))$ was seen;\\
\>\>otherwise mark it seen.\\
\>\>Compute the minimum run count $d_{\Pi'}(W_0)$ using both sources.\\
\>\>If $d_{\Pi'}(W_0)<r$, prune this extension.\\
\>\>In first mode, prune if an earlier $r$-run mask of $W_0$ exists.\\
\>\>In last mode, prune if a later $r$-run mask of $W_0$ exists.\\
\>\>In arbitrary mode, apply neither order-based pruning rule.\\
\>\>Otherwise call $\mathrm{Search}(h+1,\Pi')$.
\end{tabbing}
\end{quote}
Here $\Pi\vee\{\cdots\}$ denotes equivalence closure. The ordinary
source-grid DP computes $d_{\Pi'}(W_0)$, comparing letters by class;
retaining the lexicographically first or last path among minimum-cost
paths tests the order condition. Normalize a partition by scanning
source positions in the order $uv$ and naming classes $0,1,\ldots$ at
their first occurrence. The full memoization key is
$(L,r,M,\mathrm{mode},h,\operatorname{norm}(\Pi))$; with a fresh cache
for each fixed $(L,r,M,\mathrm{mode})$, only the last two entries
need storage. The nonmemoized mode stores no seen set and counts each
complete witness tuple separately.

At depth $h$, the partition satisfies a chosen lower witness for each
of the first $h$ moves. Every retained node still makes $M$ minimum
with $d(W_0)=r$, and first or last when that mode requires it.
Both invariants hold initially and persist after the explicit tests.
A terminal node therefore represents a mask for which every exchange
has a lower witness.

For the first-minimum-mask search, a lower or earlier pruning witness
persists under further identifications, so a pruned branch cannot yield
a counterexample whose selected mask is first and minimum. For an actual
counterexample, choose one actual lower witness for each exchange.
Every equality imposed along that witness path holds in the
counterexample. A lower or earlier original-mask pruning witness along
the path would therefore also hold in the counterexample, contradicting
its selected mask being first and minimum. The path survives, and its
leaf equality partition refines the counterexample's actual equality
pattern. Hence every counterexample is a coarsening of the partition
of a surviving leaf. This proves completeness for each tested length
and degree, over arbitrary alphabet sizes. Replacing ``earlier'' with
``later'' gives the last-mask argument; dropping that condition gives
the arbitrary-mask argument.

Memoizing the key just defined is also complete: the remaining moves,
equality constraints, and pruning tests depend only on the fixed
original mask, current partition, and depth. Earlier witness choices
with that same key have exactly the same possible continuations.

\paragraph{Execution and evidence.}
The finite premises are supplied by the completed production enumerations
archived in release v1.0, with the exact domains, commands, and completed
ranges recorded in its README. Independent degree-three symbolic
regeneration covers every counter and all twelve surviving witness tuples
through length twenty. A separate independent certificate checker verifies
all $1\,512$ archived tuples, $896$ distinct source/output/mask instances,
and $41\,244$ identification cuts through length thirty. It checks the
original degrees, lower witnesses, common periods, shifted successors,
and the persistent lower masks used below. Certificate checking establishes
the validity of the surviving cases; completeness of generation above
length twenty, and the larger higher-degree bounds, rely on the archived
completed production searches together with the reduction just proved.

\subsection{Uniform repair of degree-three bad masks}\label{sec:boundedrepair}

\begin{proposition}[Computational result]\label{prop:boundedrepair}
For combined length at most thirty, every degree-three minimum mask
whose four exchanges all have degree one has the following common-period
structure; under further identifications, the claims below are restricted
to identifications that preserve the original minimum degree three.
Writing $w=LaPbQcR$, $u=abc$, and $v=LPQR$ as in
Section~\ref{sec:periodic}, the nonempty words $P,Q,ab,bc$ are powers
of a common primitive word. The mask admits the shift
in~\eqref{eq:periodshift}, and its shifted successor-degree vector, in the
order $1L,1R,2L,2R$, is $(2,1,1,2)$, over every alphabet.
\end{proposition}

\begin{proof}
The arbitrary-mask symbolic search of
Section~\ref{sec:symbolic} represents every possible bad mask by a
coarsening of a surviving equality partition. The complete search
through length thirty yields $896$ distinct source/output/mask instances.
For each instance, the finite checks verify the original degree, all four
contiguous witnesses, the common period, and every shifted successor. The primitive periods are
$\ttw{ABA}$, $\ttw{ABCA}$, $\ttw{ABABA}$, $\ttw{ABCAB}$, and
$\ttw{ABCDA}$, with some periods represented by many distinct contexts.
These power equations persist under letter identifications, taking the
primitive root of the identified period when necessary.

Further identifications are checked without assuming that a displayed
word is the only word represented by its partition. For each possible
one-run cut of each shifted $1L$ and $2R$ successor, impose the least
additional source-letter equalities making that cut a valid mask.
The resulting original word always has degree at most two; its lower
mask remains valid under every further identification. Thus a coarsening
retaining original degree three cannot give degree one at either
shifted successor. The transported masks supply degree at most two,
while the displayed one-run witnesses for $1R$ and $2L$ persist under
identifications. This proves the vector $(2,1,1,2)$ for every such
coarsening.

The completeness argument in Section~\ref{sec:symbolic} and this persistent lower-mask
test cover every letter identification, and hence every alphabet.
\end{proof}

\subsection{All-alphabet descent bounds}

For degree three, the lower witnesses in the symbolic enumeration are
exactly the one-run masks
\[
 H_c=\V^c\U^m\V^{n-c},\qquad 0\le c\le n.
\]
In arbitrary-mask mode the search enumerates every three-run original
mask and every four-tuple of cuts, retaining the lower-run pruning
after each equality extension. A surviving tuple therefore has
original degree three and a one-run witness for each of its four
exchanges. Different tuples may describe the same equality partition
or source/output/mask instance. The completeness argument above covers
all such bad masks and every coarsening of their partitions.

\begin{proposition}[Computer-assisted]\label{prop:symbolic}
For nonempty ordered sources over any alphabet, the following hold.
\begin{enumerate}
\item Through combined length seventeen, every output of degree
$r>1$ has an exact exchange from its first minimum mask. Consequently,
Singleton No-Gap holds through length seventeen. Length eighteen is
the smallest first-minimum-mask failure over all degrees and alphabets.
\item At degree three, first-minimum-mask descent holds through
length thirty. A bad arbitrary minimum mask has combined length at
least seventeen, and Proposition~\ref{prop:family} attains that bound.
\item At degree four and length eighteen, the first-mask failure
in Proposition~\ref{prop:firstfailure} is the unique failure up to
letter renaming. At degree five, first-mask descent holds through
length eighteen. Some minimum mask has an exact exchange at each
degree $2\le r\le5$ through length eighteen. Degrees $6\le r\le9$
can occur at length eighteen, and no No-Gap assertion is made for those
degrees.
\item At degree four, last-minimum-mask descent holds through length
nineteen. At length twenty its unique failure up to letter renaming
is the example in Proposition~\ref{prop:lastfailure}, whose first
minimum mask has an exact exchange. Hence one of the two extremal
minimum masks has an exact exchange at degree four through length twenty.
\end{enumerate}
\end{proposition}
\begin{proof}
The complete symbolic searches have no surviving first-mask partition
through length seventeen, none at degree three through length thirty,
and none at degree five and length eighteen. For length at most sixteen,
the bound $d(w)\le\min(|u|,|v|+1)$ is at most eight; at length seventeen
it is at most nine. Thus all possible degrees are covered, with degree
two exact by Section~\ref{sec:unary}. Iterating exact descent proves
Singleton No-Gap in the first domain.

The arbitrary-mask degree-three search has no surviving tuple through
length sixteen. Proposition~\ref{prop:family} gives a bad mask at
length seventeen. At degree four and length eighteen the sole surviving
first-mask partition is the binary example of
Proposition~\ref{prop:firstfailure}. Its only proper coarsening is
unary and has degree one, so the failure is unique up to renaming.
Its alternative exact exchange, together with the degree-three and
degree-five searches, gives the third assertion.

The last-mask search has no degree-four survivor at length nineteen
and a single binary partition at length twenty. Its only proper
coarsening is unary, and Proposition~\ref{prop:lastfailure}, with the
complete mask list in Appendix~\ref{app:masks}, Table~\ref{tab:lastcert},
proves that its first-mask exchange is exact.
This gives the fourth assertion.

To pass from an exact-length search to shorter lengths, append $\#^q$
to $v$ and $w$, where $\#\notin\operatorname{alph}(uv)$. Every new mask
is an old mask followed by $\V^q$: an appended letter cannot belong to
$u$. Its run count and lexicographic order are unchanged, and every
exchange is an old exchange followed by $\#^q$. A shorter
counterexample would therefore persist at the tested length. A letter
different from the last letter of $u$ also forces the suffix to $\V$,
so the same padding works on any fixed alphabet with at least two letters.
\end{proof}

\subsection{A structured binary domain}

The following domain is specified separately from the exhaustive
length bounds. For each $p\in\{2,3,4\}$, take every binary block of
length $p$ containing both letters, including blocks with a smaller
period. For $3\le m,n\le12$, $m+n\le22$, and $0\le f<p$, let $u$
be the length-$m$ prefix of its infinite repetition starting at offset
zero, and $v$ the length-$n$ prefix starting at offset $f$. When
$m+n\le18$, also include each single-position flip of $u$ with $v$
fixed, and each single-position flip of $v$ with $u$ fixed; never
perturb both sources. Deduplicate the exact ordered pairs. Denote the
resulting set of $70\,710$ pairs by $\mathcal S$.

\begin{proposition}[Computer-assisted]\label{prop:sample}
For every $(u,v)\in\mathcal S$, every output of degree $r>1$ has an
exact exchange from each extremal minimum mask. Every degree-three
minimum mask has at least one exact exchange.
\end{proposition}

The grouped-mask enumeration verifies the extremal assertions. A
separate pass through all degree-three output groups examines every
minimum mask and verifies the second assertion. The domain contains
$20\,001\,166$ outputs, including $8\,861\,068$ of degree three;
this conclusion is specific to $\mathcal S$.

\subsection{The prefix test}

An eligible adjacent swap interchanges unequal letters at positions
$i,i+1$ assigned to different sources by some minimum mask, whose
transported mask has one fewer run. Count positions once within a
source/output triple, even if several masks make them eligible.

\begin{proposition}[Computer-assisted]\label{prop:profiles}
Among all labeled binary nonempty ordered source pairs of combined
length eight or nine and their degree-three outputs, exactly $168$
triples $(u,v,w)$ have an eligible swap whose output has degree one.
Each triple has exactly one such failed position. Of the $116$ triples
having another eligible swap of degree two, $94$ have such a swap
certified by~\eqref{eq:profilebound}, and $22$ have none certified.
For each of the remaining $52$ triples, the prefix test certifies an
ordinary adjacent swap of degree two, which need not lower a minimum
mask's run count.
\end{proposition}

Enumeration of every mask and adjacent position verifies the counts;
the prefix dynamic program tests whether there exists a cutoff
$i+2\le t<|w|$ satisfying~\eqref{eq:profilebound} at every source state.
The same word from a different ordered source pair is counted separately.
The lower-bound argument in Section~\ref{sec:profiles} proves the
meaning of each certificate. Thus the prefix condition is useful for
certifying individual predecessors but does not characterize existence
of a successful eligible swap.

\section{Iteration depth for arbitrary languages}\label{sec:depth}

Hughes proved iteration-depth No-Gap for finite
languages~\cite[Section~10, Theorem~8]{hughes} and posed the
Iteration-Depth No-Gap Conjecture for arbitrary
languages~\cite[Section~10]{hughes}. The following padding argument
proves that conjecture.

Let $A,B\subseteq\Sigma^*$ be arbitrary languages. For a positive
integer $k$, a $k$-insertion writes a source word as
$y_1\cdots y_k\in A$ and a base word as $x_1\cdots x_{k+1}$,
and produces $x_1y_1\cdots x_ky_kx_{k+1}$. Factors may be empty.
Set $L_{k,0}=B$ and obtain $L_{k,t+1}$ by $k$-insertion into a base
word in $L_{k,t}$. The degree $k$ is fixed within each history.
For a reachable word define
\[
 \delta(w)=\min\{t\ge0:\exists k\ge1,\ w\in L_{k,t}\}.
\]

\begin{theorem}\label{thm:depth}
For arbitrary languages $A,B$, if a reachable word $w$ has
$\delta(w)=r$, then for each integer $q$ with $0\le q\le r$
there exists a reachable word $w_q$ with $\delta(w_q)=q$.
\end{theorem}
\begin{proof}
Padding by empty source and base factors gives
$L_{j,t}\subseteq L_{K,t}$ whenever $1\le j\le K$; induction on $t$
lifts the predecessor and then pads the last insertion.
Suppose $w\in L_{k,r+1}$ with $\delta(w)=r+1$, and take its final
base predecessor $x\in L_{k,r}$. If $\delta(x)=s<r$ through degree
$j$, let $K=\max(j,k)$. Padding gives $x\in L_{K,s}$, and padding
the final insertion gives $w\in L_{K,s+1}$. This contradicts minimality
of $r+1$. Therefore $\delta(x)=r$, and induction realizes all smaller
depths. No inclusion between successive iteration stages is required.
\end{proof}

The iteration-depth theorem above has a Lean~4 formalization~\cite{lean}
for arbitrary languages over an arbitrary alphabet type
$\alpha:\mathrm{Type}$, without finiteness or countability assumptions.
The formal theorem \texttt{result} uses the literal insertion operation
with one fixed positive degree throughout each history. Its empty-pair
padding argument synchronizes degrees and shows that a predecessor of a
word of minimum depth $r+1$ has minimum depth exactly $r$.

\section{Conclusion and open problems}\label{sec:conclusion}

Block exchange has an exact algebraic obstruction: the commuting
representations of the successor have minimum degree $r-1$, while
every additional decrease is caused by a noncommuting representation.
The unary formulas eliminate this obstruction for every minimum mask.
In the periodic degree-three family, an explicit label shift yields
an exact exchange. The counterexamples show why a general descent
argument must allow more than an arbitrary mask, either extremal mask,
or a fixed separator rule.

The principal open question is whether every shuffle of degree $r>1$
admits some minimum mask and some block exchange with no noncommuting
representation of at most $r-2$ runs. An affirmative answer would prove
Singleton No-Gap. The finite results establish this conclusion through
length seventeen in all degrees, and give larger bounds at selected
degrees, but the unbounded existence step remains unresolved.

At degree three, two structural questions are more specific. Must every
bad minimum mask have $P,Q,ab,bc$ as powers of one primitive word?
Under that hypothesis and arbitrary outer contexts, must the shift
in~\eqref{eq:periodshift} have an exact $1L$ exchange?
Theorem~\ref{thm:periodrepair} answers the repair question for its
infinite family; Proposition~\ref{prop:boundedrepair} answers both
questions through length thirty. Establishing the common-period
classification and extending the repair to all contexts would give
a route to an unbounded degree-three theorem.

At the language level, Hughes's representation-erasure and
finite-to-arbitrary questions remain~\cite[Section~11]{hughes}.
Even if every singleton spectrum is an initial interval, competing
source pairs can lower degrees of the same output, so the language
spectrum requires a separate argument. Iteration depth has a different
mechanism: padding all histories to a common insertion degree makes
the last predecessor attain the preceding minimum depth. That proves
No-Gap for arbitrary languages without an assumption of monotonicity
between successive iteration stages.

\clearpage
\appendix
\section{Complete finite mask lists}\label{app:masks}

The tables below retain every mask used in the finite completeness
arguments of Section~\ref{sec:examples}. For fixed $u,v,w$, a direct
position test enumerates every $|u|$-element subset of the positions of
$w$, retains it exactly when its subsequence is $u$ and the complementary
subsequence is $v$, and counts its maximal $\U$-runs. Thus the lists
include every representation for the specified ordered sources.

Table~\ref{tab:masks} gives the sixteen masks in
Proposition~\ref{prop:badmask}; its proof also derives the list directly
from the four equal-letter pairs.

\begin{table}[ht]
\centering\small
\caption{All masks of $\ttw{0010\hspace{0.25em}1100\hspace{0.25em}1101\hspace{0.25em}0}$ for the sources in
Proposition~\ref{prop:badmask}. Each row pairs the two possibilities for
the initial two positions.}\label{tab:masks}
\begin{tabular}{@{}lclc@{}}
\toprule Mask & Runs & Mask & Runs\\\midrule
\ttw{UVVVUVUVUVVVU}&5&\ttw{VUVVUVUVUVVVU}&5\\
\ttw{UVVVUVUVVUVVU}&5&\ttw{VUVVUVUVVUVVU}&5\\
\ttw{UVVVUVVUUVVVU}&4&\ttw{VUVVUVVUUVVVU}&4\\
\ttw{UVVVUVVUVUVVU}&5&\ttw{VUVVUVVUVUVVU}&5\\
\ttw{UVVVVUUVUVVVU}&4&\ttw{VUVVVUUVUVVVU}&4\\
\ttw{UVVVVUUVVUVVU}&4&\ttw{VUVVVUUVVUVVU}&4\\
\ttw{UVVVVUVUUVVVU}&4&\ttw{VUVVVUVUUVVVU}&4\\
\ttw{UVVVVUVUVUVVU}&5&\ttw{VUVVVUVUVUVVU}&5\\
\bottomrule
\end{tabular}
\end{table}

Table~\ref{tab:lastcert} gives all masks of the original word and the
exact first-mask successor in Proposition~\ref{prop:lastfailure}.
The minima in the two lists are four and three, respectively.

\begin{table}[ht]
\centering\footnotesize
\caption{Complete mask lists for the last-mask example and its exact
first-mask successor $z_*$. The upper list has eight masks; the lower
list has sixteen. Both use the same ordered sources.}\label{tab:lastcert}
\textit{Original word $w$}\par\smallskip
\begin{tabular}{@{}lrlr@{}}
\toprule Mask & $\rho$ & Mask & $\rho$\\\midrule
\ttw{UUUUUUUVVUVUVVVUVVUV} & 5 & \ttw{UUUUUUUVVUVUVVVUVVVU} & 5\\
\ttw{UUUUUUUVVUVVUVVUVVUV} & 5 & \ttw{UUUUUUUVVUVVUVVUVVVU} & 5\\
\ttw{UUUUUUUVVVUUVVVUVVUV} & 4 & \ttw{UUUUUUUVVVUUVVVUVVVU} & 4\\
\ttw{UUUUUUUVVVUVUVVUVVUV} & 5 & \ttw{UUUUUUUVVVUVUVVUVVVU} & 5\\
\bottomrule
\end{tabular}
\par\medskip
\textit{Exact successor $z_*=\ttw{ABABABBABAABBABBAABB}$}\par\smallskip
\begin{tabular}{@{}lrlr@{}}
\toprule Mask & $\rho$ & Mask & $\rho$\\\midrule
\ttw{UUUUUUUVVUVUVVVVUVUV} & 5 & \ttw{UUUUUUUVVUVUVVVVUVVU} & 5\\
\ttw{UUUUUUUVVUVUVVVVVUUV} & 4 & \ttw{UUUUUUUVVUVUVVVVVUVU} & 5\\
\ttw{UUUUUUUVVUVVUVVVUVUV} & 5 & \ttw{UUUUUUUVVUVVUVVVUVVU} & 5\\
\ttw{UUUUUUUVVUVVUVVVVUUV} & 4 & \ttw{UUUUUUUVVUVVUVVVVUVU} & 5\\
\ttw{UUUUUUUVVVUUVVVVUVUV} & 4 & \ttw{UUUUUUUVVVUUVVVVUVVU} & 4\\
\ttw{UUUUUUUVVVUUVVVVVUUV} & 3 & \ttw{UUUUUUUVVVUUVVVVVUVU} & 4\\
\ttw{UUUUUUUVVVUVUVVVUVUV} & 5 & \ttw{UUUUUUUVVVUVUVVVUVVU} & 5\\
\ttw{UUUUUUUVVVUVUVVVVUUV} & 4 & \ttw{UUUUUUUVVVUVUVVVVUVU} & 5\\
\bottomrule
\end{tabular}
\end{table}

\clearpage
Table~\ref{tab:packedcert} gives all masks of the packed example and
its first-mask successor from Section~\ref{sec:examples}. Their minima
are four and three, respectively.

\begin{table}[ht]
\centering\footnotesize
\caption{Complete mask lists for the packed example: eight masks of
$w$, and twenty-one masks of the exact first-mask successor
$z_\dagger$.}\label{tab:packedcert}
\textit{Original word $w$}\par\smallskip
\begin{tabular}{@{}lrlr@{}}
\toprule Mask & $\rho$ & Mask & $\rho$\\\midrule
\ttw{UUUVVUVUUVUUVVVVV} & 4 & \ttw{UUUVVUVUUVVVUUVVV} & 4\\
\ttw{UUUVVUVVVUUVUUVVV} & 4 & \ttw{UUUVVUVVVVVUVVUUU} & 4\\
\ttw{VVVUUVUUVUUVUUVVV} & 4 & \ttw{VVVUUVUUVVVUVVUUU} & 4\\
\ttw{VVVUUVVVUUVUVVUUU} & 4 & \ttw{VVVVVUUVUUVUVVUUU} & 4\\
\bottomrule
\end{tabular}
\par\medskip
\textit{Exact successor $z_\dagger=\ttw{ABABBABABABABABBA}$}\par\smallskip
\begin{tabular}{@{}lrlrlr@{}}
\toprule Mask & $\rho$ & Mask & $\rho$ & Mask & $\rho$\\\midrule
\ttw{UUVUVUVUUVUUVVVVV} & 5 & \ttw{UUVUVUVUUVVVUUVVV} & 5 & \ttw{UUVUVUVVVUUVUUVVV} & 5\\
\ttw{UUVUVUVVVVVUVVUUU} & 5 & \ttw{UUVVUUVUUVUUVVVVV} & 4 & \ttw{UUVVUUVUUVVVUUVVV} & 4\\
\ttw{UUVVUUVVVUUVUUVVV} & 4 & \ttw{UUVVUUVVVVVUVVUUU} & 4 & \ttw{UUVVVVUUVUUVUUVVV} & 4\\
\ttw{UUVVVVUUVVVUVVUUU} & 4 & \ttw{UUVVVVVVUUVUVVUUU} & 4 & \ttw{VVUUUUVUUVUUVVVVV} & 3\\
\ttw{VVUUUUVUUVVVUUVVV} & 3 & \ttw{VVUUUUVVVUUVUUVVV} & 3 & \ttw{VVUUUUVVVVVUVVUUU} & 3\\
\ttw{VVUUVVUUVUUVUUVVV} & 4 & \ttw{VVUUVVUUVVVUVVUUU} & 4 & \ttw{VVUUVVVVUUVUVVUUU} & 4\\
\ttw{VVUVUVUUVUUVUUVVV} & 5 & \ttw{VVUVUVUUVVVUVVUUU} & 5 & \ttw{VVUVUVVVUUVUVVUUU} & 5\\
\bottomrule
\end{tabular}
\end{table}

\clearpage
\section*{Use of AI tools}
The authors used AI tools throughout this work. Ma and Zhang used ChatGPT Pro (GPT-6 Pro), Codex (gpt-6.1-sol) and Claude Opus 5.5 (via Claude Code) for literature search, computation and double-checking, including formal verification of the results in Lean; Hughes used ChatGPT for mathematical discussions and the development of open problems, for critical examination of arguments and for suggestions on exposition and manuscript organization, and Claude.ai to validate all proofs and computations. Ma and Zhang chose the open questions on which this paper is based and developed the initial proofs and counterexamples, which all three authors then refined together. All proofs can be checked by hand, the computations can be reproduced from the verification package, and all authors take full responsibility for the final manuscript.

\end{document}